\documentclass[aps,showpacs,preprintnumbers,amsmath, amssymb]{revtex4}

\usepackage{float}
\usepackage{graphics,epsfig}
\usepackage{graphicx}
\usepackage{dcolumn}
\usepackage{bm}

\begin{document}
\baselineskip=0.8 cm

\title{{\bf Holographic superconductor models in the non-minimal derivative coupling theory}}

\author{Songbai Chen$^{1}$\footnote{Email: csb3752@163.com}, Qiyuan Pan$^{1,2}$\footnote{Email: panqiyuan@126.com},
 Jiliang Jing $^{1}$\footnote{Email: jljing@hunnu.edu.cn}}
\affiliation{$^{1}$ Institute of Physics and Department of Physics,
Hunan Normal University,  Changsha, Hunan 410081, P. R. China \\ Key Laboratory of Low Dimensional Quantum Structures \\
and Quantum Control of Ministry of Education, Hunan Normal
University, Changsha, Hunan 410081, P. R. China} \affiliation{$^{2}$
Department of Physics, Fudan University, Shanghai 200433, P. R.
China.}

\vspace*{0.2cm}
\begin{abstract}
\baselineskip=0.6 cm
\begin{center}
{\bf Abstract}
\end{center}

We study a general class of holographic superconductor models via
the St\"{u}ckelberg mechanism in the non-minimal derivative coupling
theory in which the charged scalar field is kinetically coupling to
Einstein's tensor. We explore the effects of the coupling parameter
on the critical temperature, the order of phase transitions and the
critical exponents near the second-order phase transition point.
Moreover, we compute the electric conductive using the probe
approximation and check the ratios $\omega_g/T_c$ for the different
coupling parameters.

\end{abstract}

 \maketitle
\newpage
\section{Introduction}

One of the most significant discovery in the string theory is the
AdS/CFT correspondence \cite{ads1,ads2,ads3} which relates a weakly
coupled gravity theory in $d+1$-dimensional AdS spacetime to a
strongly coupled conformal field theory living on the
$d$-dimensional boundary. It is widely believed now that the AdS/CFT
correspondence could provide a new means of studying the strongly
interacting condensed matter systems in which the perturbative
methods are no longer valid. Thus, this holographic method is
expected to give us some insights into the nature of the pairing
mechanism in the high temperature superconductors which fall outside
of the scope of current theories of superconductivity. In the
pioneering papers \cite{Hs01,Hs0,a1,Hs02,Hs03},  it was suggested
that the spontaneous $U(1)$ symmetry breaking happened in the bulk
black holes can be used to build gravitational duals of the
transition from normal state to superconducting state in the
boundary theory. This dual model consists of a system with a black
hole and a charged scalar field, in which the black hole admits
scalar hair at temperature $T$ smaller than a critical temperature
$T_c$. The appearance of a hairy AdS black hole means the formation
of the scalar condensation in the boundary CFT, which indicates that
the expectation value of charged operators undergoes the U(1)
symmetry breaking and then the phase transition is occurred. It was
argued that this phase transition belongs to the second order. Due
its potential applications to the condensed matter physics,
holographic superconductors in the various theories of gravity have
been investigated  in the recent years
\cite{Hsf0,Hsf01,Hsa1,a0,a01,a2,a3,a4,a401,a40,a402,a41,a42,
a5,a6,a7,a8,a8d1,GT,a81,a9,a10,a100,a20,a21,a22,a23,a13,a14,a15,a151,a152,a16s,a17}.
Recently, Franco \textit{et al.} \cite{a18,a19} investigated the
general models of holographic superconductors in which the
spontaneous breaking of a global U(1) symmetry occurs via the
St\"{u}ckelberg mechanism. They found that both the order of the
phase transition and the critical exponents for the second phase
transition can be tuned by the parameters defined in the models.
These generalized holographic superconductor models have also been
extended recently to the Gauss-Bonnet gravity \cite{ap1} and
Born-Infeld electrodynamics \cite{ap2}.

It is well known that the properties of holographic superconductors
depend on the coupling between the scalar field and the
electromagnetic field. Theoretically, the general form of the action
with more couplings can be expressed as
\begin{eqnarray}
S=\int d^4x \sqrt{-g}\bigg[f(\psi, R, R_{\mu\nu}R^{\mu\nu},
R_{\mu\nu\rho\sigma}R^{\mu\nu\rho\sigma}) +K(\psi,
\partial_{\mu}\psi\partial^{\mu}\psi,\nabla^2\psi, R^{\mu\nu}\partial_{\mu}\psi\partial_{\nu}\psi,\cdot\cdot\cdot)+V(\psi)+Y(\psi)\mathcal{L}_m\bigg],\label{act2}
\end{eqnarray}
where $f$, $K$ and $Y$ are arbitrary functions of the corresponding
variables, $\mathcal{L}_m$ is the Lagrangian for other matter
fields. Obviously, the nonlinear functions $f$, $K$ and $Y$ provide
the more non-minimal couplings among fields and the background
spacetime. These new couplings modify the usual Klein-Gordon
equation and Einstein-Maxwell equation so that the motion equations
for the scalar field and electromagnet filed are no longer generally
the second-order differential equations, which may changes the
properties of holographic superconductors in the AdS spacetime.
Recently, F. Aprile \textit{et al.} \cite{a12} considered a special
case of the action (\ref{act2}) in which the system contains the
couplings among a scalar field, electromagnetic field and the
cosmological constant so that the action has a form
\begin{eqnarray}
S=\int d^4x
\sqrt{-g}\bigg[R-\frac{1}{4}G(\psi)F^{\mu\nu}F_{\mu\nu}+\frac{6}{L^2}U(\psi)-\frac{1}{2}\partial_{\mu}\psi\partial^{\mu}\psi
-\frac{1}{2}J(\psi)A_{\mu}A^{\mu}\bigg].
\end{eqnarray}
They studied the properties of holographic superconductors and found
that some aspects of the quantum critical behavior strongly depend
on the choice of couplings $G(\psi)$, $U(\psi)$ and $J(\psi)$.
Another interesting case is that the scalar field in the action
(\ref{act2}) is kinetically coupling with Einstein's tensor. Sushkov
found \cite{Sushkov:2009} that the equation of motion for the scalar
field can be reduced to second-order differential equation in this
model, which means that the theory is a ``good" dynamical theory
from the point of view of physics. The applications of this model to
the cosmology has been done extensively in
\cite{Sushkov:2009,Gao:2010, Granda:2009,Saridakis:2010}. It is
found that in cosmology the problem of graceful exit from inflation
with such a coupling  has a natural solution without any fine-tuned
potential. The dynamical behaviors and Hawking radiation of a scalar
field coupling to Einstein's tensor in the background of a black
hole spacetime have been studied in \cite{sc1,Chen:2010,sc1}. It was
showed that this coupling changes the stability of the black hole
and enhances Hawking radiation of the black hole.  The main purpose
of this paper is to investigate the properties of holographic
superconductors in the models in which a charged scalar field is
kinetically coupling to Einstein's tensor and probe the effects of
such a coupling on physical quantities and on the dynamics of the
phase transition.

This paper is organized as follows. In Sec. II, we will study the
scalar condensation and the phase transitions for a general class of
the holographic superconductor models via the St\"{u}ckelberg
mechanism as the charged scalar field is kinetically coupling to
Einstein's tensor. In Sec. III, we will probe the effects of the
coupling parameter and other model parameters on the conductivity of
the charged condensates. Finally, we will conclude in the last
section of our main results.

\section{General holographic superconductor models in the non-minimal derivative coupling theory}
We will consider the action of the scalar field coupling to the
Einstein's tensor $G^{\mu\nu}$ in the AdS background
\begin{eqnarray}
S=\int d^4x
\sqrt{-g}\bigg[R-\Lambda-\frac{F^2}{4}-\frac{|D_{\mu}\tilde{\psi}D^{\mu}\tilde{\psi}|}{2}-
\frac{b |G^{\mu\nu}D_{\mu}\tilde{\psi}D_{\nu}\tilde{\psi}|}{2}-
\frac{m^2|\tilde{\psi}^2|}{2}-V(|\tilde{\psi}|)\bigg],\label{acts}
\end{eqnarray}
where $D_{\mu}=\partial_{\mu}-iA_{\mu}$ and $b$ is the coupling
parameter. As in \cite{a18,a19}, re-scaling the charged scalar field
$\tilde{\psi}=\psi e^{ip}$ with real $\psi$ and $p$, and then the
action (\ref{acts}) can be re-written in a St\"{u}ckelberg form
\begin{eqnarray}
S&=&\int d^4x
\sqrt{-g}\bigg[R-\Lambda-\frac{F^2}{4}-\frac{(g^{\mu\nu}+bG^{\mu\nu})\partial_{\mu}\psi\partial_{\nu}\psi}{2}\nonumber\\&-&
\frac{m^2\psi^2}{2}-\frac{\psi^2(g^{\mu\nu}+bG^{\mu\nu})(\partial_{\mu}p-A_{\mu})(\partial_{\nu}p-A_{\nu})}{2}-V(\psi)\bigg],\label{acts1}
\end{eqnarray}
with the gauge symmetry $A_{\mu}\rightarrow A_{\mu}+\partial_{\mu}
\alpha$ together with $p\rightarrow p+\alpha$. Thus, the generalized
St\"{u}ckelberg Lagrangian action in the non-minimal derivative
coupling theory can be expressed as
\begin{eqnarray}
S&=&\int d^4x
\sqrt{-g}\bigg[R-\Lambda-\frac{F^2}{4}-\frac{(g^{\mu\nu}+bG^{\mu\nu})\partial_{\mu}\psi\partial_{\nu}\psi}{2}\nonumber\\&-&
\frac{m^2\psi^2}{2}-\frac{|\mathfrak{F}(\psi)|(g^{\mu\nu}+bG^{\mu\nu})(\partial_{\mu}p-A_{\mu})(\partial_{\nu}p-A_{\nu})}{2}-V(\psi)\bigg].\label{acts2}
\end{eqnarray}
Here $\mathfrak{F}(\psi)$ is a function of $\psi$, which is take as
\begin{eqnarray}
\mathfrak{F}(\psi)=\psi^2+c_{\gamma}\psi^{\gamma}+c_4\psi^4,
\end{eqnarray}
with the model parameters $c_{\gamma}$,  $\gamma$ and $c_4$
\cite{a18,a19,ap1,ap2}. In the probe limit, the backreactions on the
background metric from the matter field is negligible.  Here we
adopt to such a limit  so that we can neglect the influence of the
coupling terms contained the factor $bG^{\mu\nu}$ in the action
(\ref{acts2}) on the metric for simplicity.

One of the most simple AdS black hole is the planar Schwarzschild
AdS black hole which can be described by a metric
\begin{eqnarray}
ds^2=-f(r)dt^2+\frac{1}{f(r)}dr^2+r^2(dx^2+dy^2),\label{m1}
\end{eqnarray}
with
\begin{eqnarray}
f(r)=\frac{r^2}{L^2}-\frac{2M}{r},
\end{eqnarray}
where $L$ is the radius of AdS and $M$ is the mass of black hole.
The Hawking temperature is
\begin{eqnarray}
T_H=\frac{3r_H}{4\pi L^2},
\end{eqnarray}
where $r_H$ is the event horizon of the black hole. The Einstein's
tensor $G^{\mu\nu}$ in the Schwarzschild AdS black hole (\ref{m1})
has a form
\begin{eqnarray}
G^{\mu\nu}=\frac{3}{L^2}g^{\mu\nu}.\label{Ein2}
\end{eqnarray}
Substituting Eqs. (\ref{m1}) and (\ref{Ein2}) into Eq. (\ref{acts2})
and taking the ansatz
\begin{eqnarray}
A_{\mu}=(\phi(r),0,0,0),\;\;\;\;\psi=\psi(r),
\end{eqnarray}
we can find that the equations of motion for the complex scalar
field $\psi$ and electrical scalar potential $\phi(r)$ can be
written as
\begin{eqnarray}
\psi^{''}+(\frac{f'}{f}+\frac{2}{r})\psi'
+\frac{\phi^2}{2f^2}\frac{d\mathfrak{F(\psi)}}{d\psi}-\frac{m^2L^4\psi}{(L^2+3b)f}=0,\label{emo1}
\end{eqnarray}
and
\begin{eqnarray}
\phi^{''}+\frac{2}{r}\phi'
-(1+3b)\frac{\mathfrak{F(\psi)}}{f}\phi=0,\label{emo2}
\end{eqnarray}
respectively. Here a prime denotes the derivative with respect to
$r$ and we use the gauge freedom to fix $p=0$. Obviously, the
presence of the coupling factor $b$ decreases the effective mass
$m^2_{eff}=\frac{m^2L^2}{L^2+3b}$ of the scalar field and increases
the current of the electric field. This means that the coupling term
will change the properties of holographic superconductors in the
Schwarzschild AdS black hole. In order to solve the nonlinear
equations (\ref{emo1}) and (\ref{emo2}) numerically, we need to seek
the boundary condition for $\phi$ and $\psi$ near the black hole
horizon $r\sim r_H$ and at the asymptotic AdS boundary
$r\rightarrow\infty$. The regularity condition at the horizon gives
the boundary conditions $\phi(r_H)=0$ and
$\psi=\frac{3(L^2+3b)r_H}{m^2L^4}\psi'$. At the asymptotic AdS
boundary $r\rightarrow\infty$, the scalar filed $\psi$ and the
electric potential $\phi$ can be approximated as
\begin{eqnarray}
\psi=\frac{\psi_{-}}{r^{\lambda_{-}}}+\frac{\psi_{+}}{r^{\lambda_{+}}},\label{b1}
\end{eqnarray}
and
\begin{eqnarray}
\phi=\mu-\frac{\rho}{r},\label{b2}
\end{eqnarray}
with
\begin{eqnarray}
\lambda_{\pm}=\frac{3}{2}\pm\sqrt{\frac{9}{4}+\frac{m^2L^4}{L^2+3b}}.
\end{eqnarray}
In the dual field theory, the constants $\mu$ and $\rho$ are
interpreted as the chemical potential and charge density
respectively. The coefficients $\psi_{-}$ and $\psi_{+}$ correspond
to the vacuum expectation values of the condensate operators
$\mathcal{O_{\pm}}$ respectively. Since both of the falloffs are
normalizable for $\psi$, one can impose the boundary condition that
either $\psi_{-}$ or $\psi_{+}$ vanishes, which ensures that the
theory is stable in the asymptotic AdS region \cite{ads}. The
choices of the boundary condition $\psi_{-}=0$ or $\psi_{+}=0$ will
not affect qualitatively our results. Thus, we here set
$\psi_{-}=0$, $L=1$ and $m^2L^2=-2$ for convenience. In doing so, we
can investigate the properties of the scalar condensate
$\langle\mathcal{O_{+}}\rangle=\psi_{+}$ for fixed charge density by
solving the equations of motion (\ref{emo1}) and (\ref{emo2})
numerically.

Let us now to study how the phase transition depends on the coupling
parameter $b$ and the model coefficients $c_{\gamma}$, $\gamma$ and
$c_4$ in the $\mathfrak{F(\psi)}$. In Fig. 1, we present the
influence of the parameters $b$ and $c_4$ on the phase transition
for fixed $c_{\gamma}=0$. We find that the coupling parameter $b$
and the model parameters $c_4$ have obvious different effects on the
critical temperature. In the case $c_4<1$, we find that only the
second-order phase transition happens in the formation of scalar
hair and the critical temperature does not depend on the coefficient
$c_4$. These properties of Holographic superconductors are similar
to that in the case in which the scalar field is not coupling with
Einstein's tensor. However, with the increase of the coupling
parameter $b$, the critical temperature decreases, which means that
the coupling between the scalar field and Einstein's tensor makes
the formation of the scalar condensate more hardly. This property of
the coupling parameter $b$ has not been observed elsewhere. For the
$c_4\geq1$, it was argued that for the case $b=0$ the behavior of
the scalar condensate means that the phase transition changes from
the second order to the first order. Here we choose $c_4=1.2, 1.4,
1.6$ and $2.0$, and probe the effect of $b$ on the transition point
of the phase transition from the second order to the first order.
Fig. 1 tells us that for the bigger value of $b$, the transition
point appears more hardly. The critical value of $b_c$ separating
the second order and the first order phase transitions is listed in
table I for selected $c_4$ within the range $[1.2, 2.0]$. Comparing
with the discussions in \cite{ap1}, one can find that the influences
of $b$ on the phase transition are different from that of the
Gauss-Bonnet coupling parameter $\alpha$.
\begin{table}[ht]
\caption{The critical value of $b_c$ which can separate the first-
and second-order behavior for different
$\mathfrak{F}(\psi)=\psi^2+c_4\psi^4$.}
\begin{tabular}[b]{cccccc}
\hline\hline
 $c_4$&$1.2$&$1.4$&$1.6$
&$2.0$\\
 $ b_c$&$0.1$&$0.2$&$0.3$&
   $0.5$\\
 \hline\hline
\end{tabular}
\end{table}
\begin{figure}[ht]
\begin{center}
\includegraphics[width=7cm]{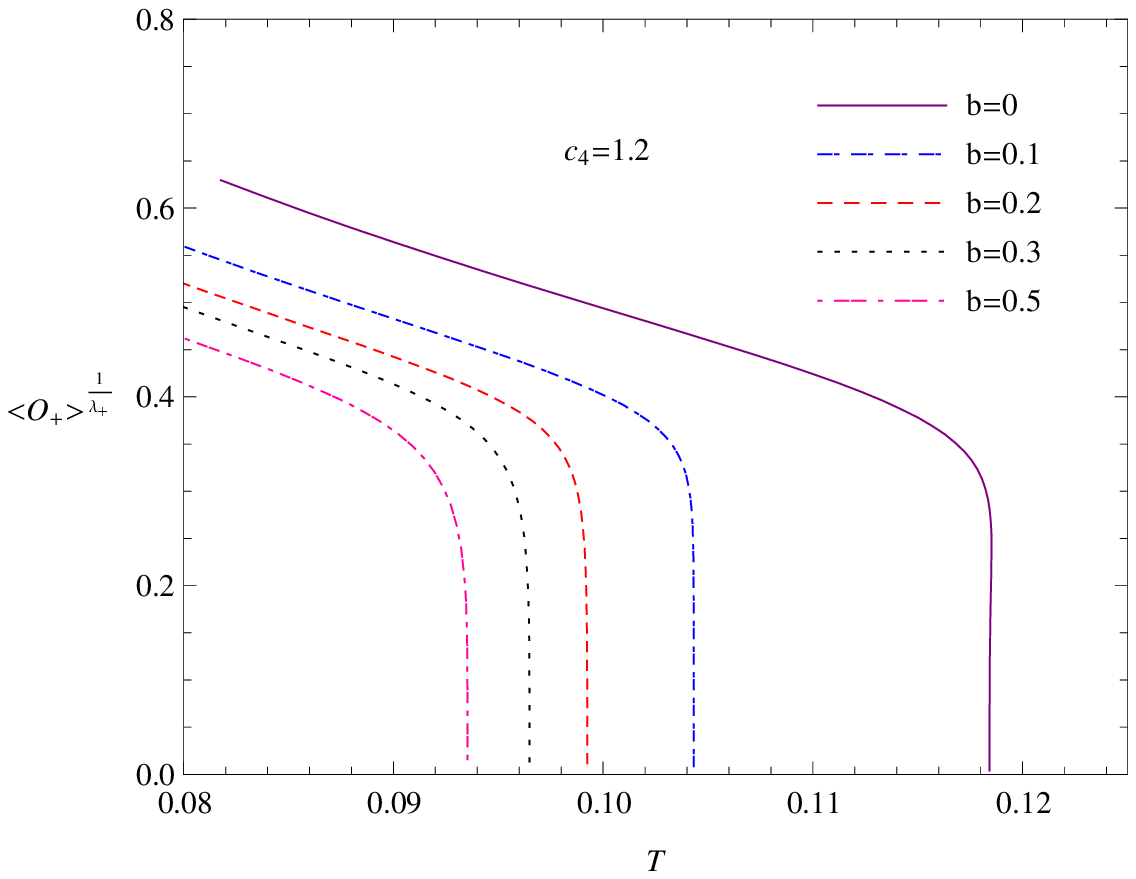}\;\;\;\;
\includegraphics[width=7cm]{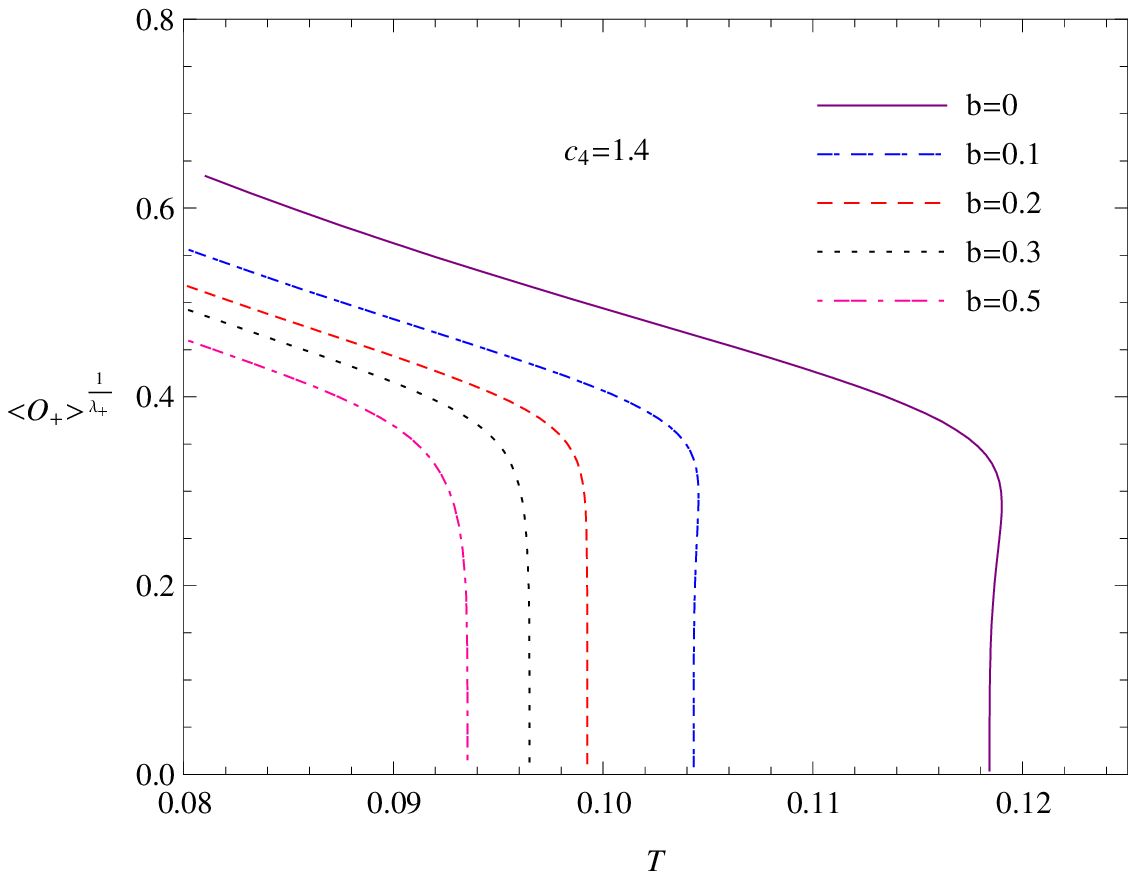}\\
\includegraphics[width=7cm]{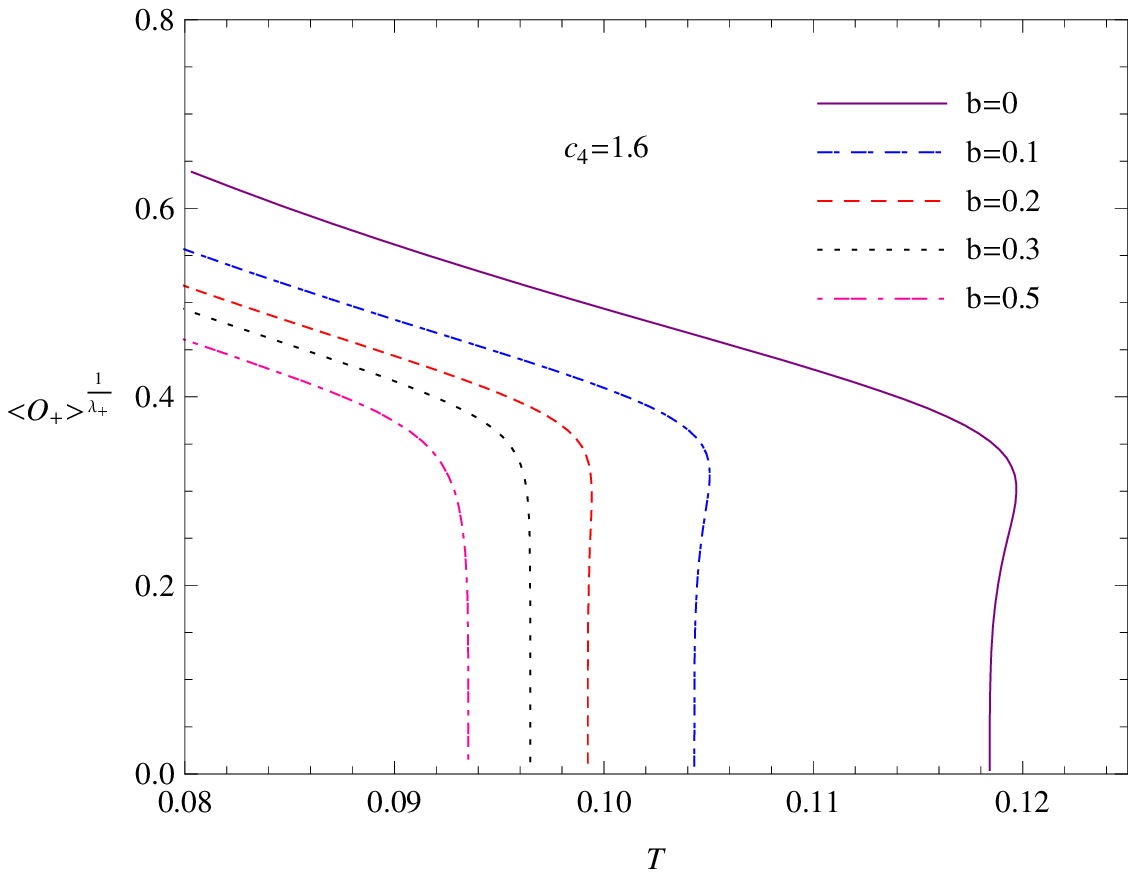}\;\;\;\;\includegraphics[width=7cm]{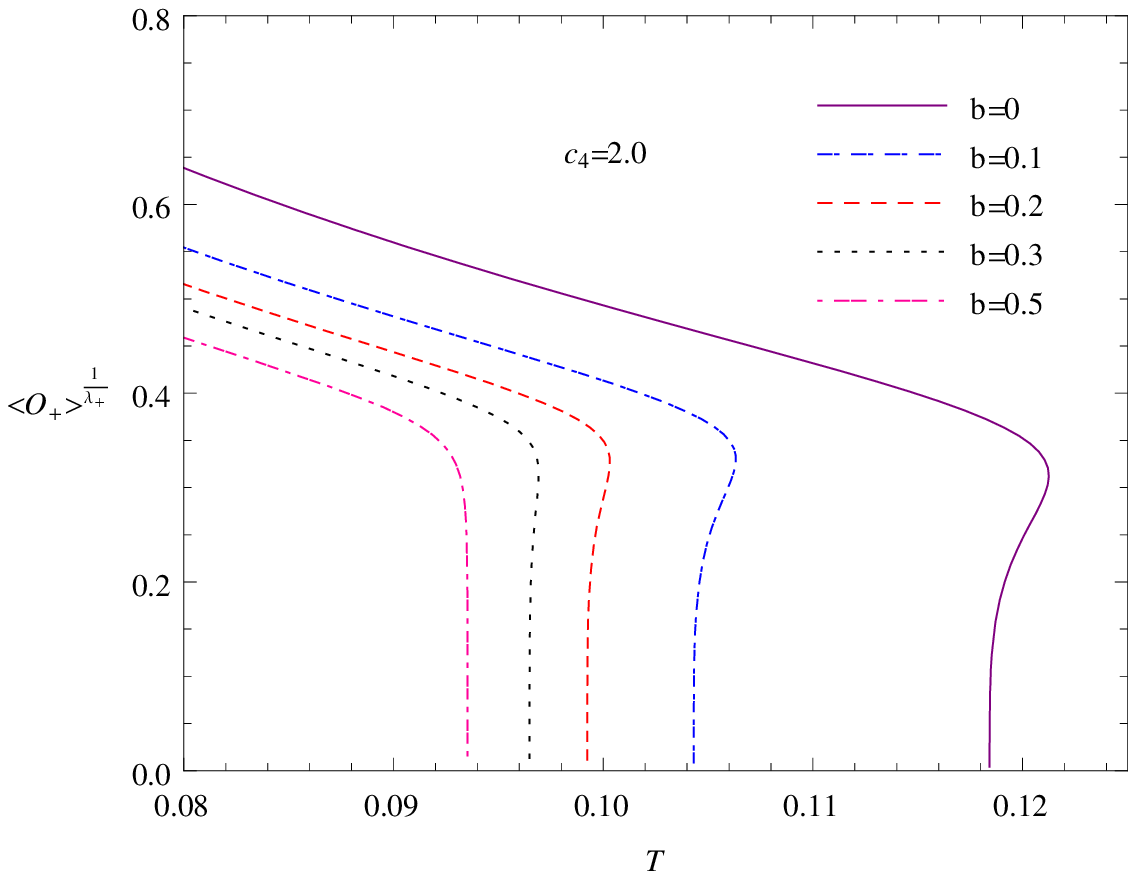}
\caption{The condensate
$\langle\mathcal{O_{+}}^{\frac{1}{\lambda_+}}\rangle$ as a function
of temperature with fixed values $c_{\gamma}=0$ for different values
of $b$ and  $c_4$, which shows that the different values of these
parameters not only change the formation of the scalar hair, but
also affect the separation between the first- and second-order phase
transitions. The five lines in each panel from left to right
correspond to decreasing $b$, i.e, 0.5, 0.3, 0.2, 0.1 and 0.}
\end{center}
\end{figure}

We also consider the case $c_{\gamma}\neq0$ for the model
$\mathfrak{F}(\psi)=\psi^2+c_{\gamma}\psi^{\gamma}+c_4\psi^4$. From
Fig. 2, we find that the critical temperature increases with the
\begin{figure}[ht]
\begin{center}
\includegraphics[width=7cm]{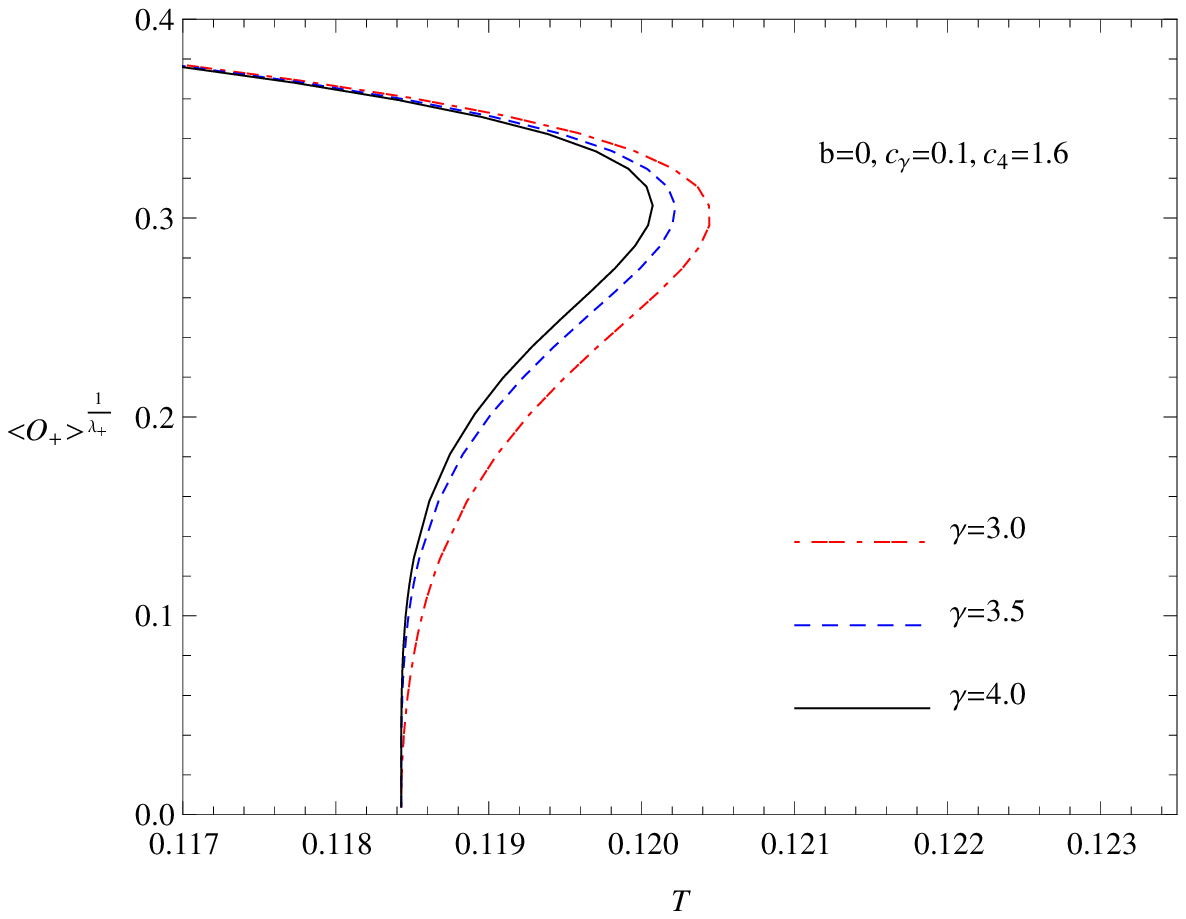}\;\;\;\;
\includegraphics[width=7cm]{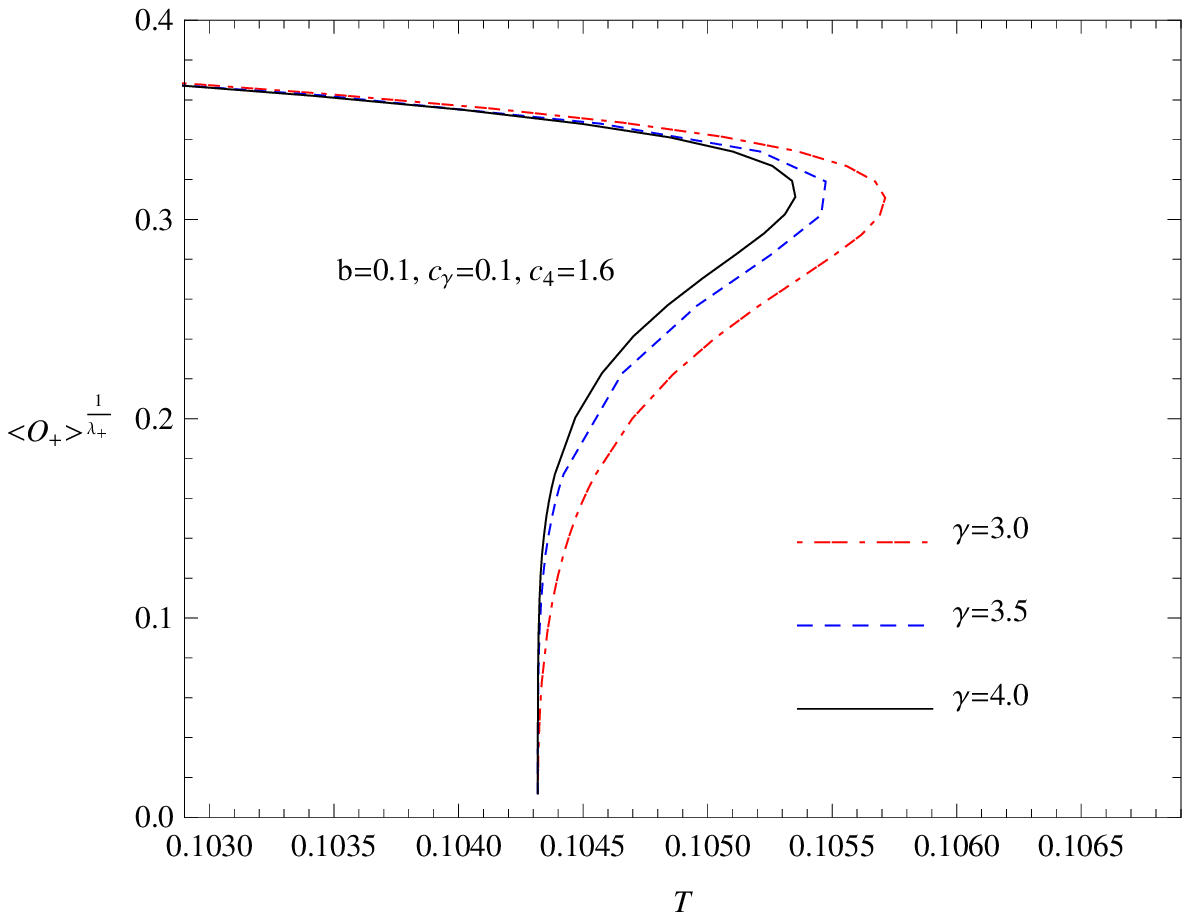}\\
\includegraphics[width=7cm]{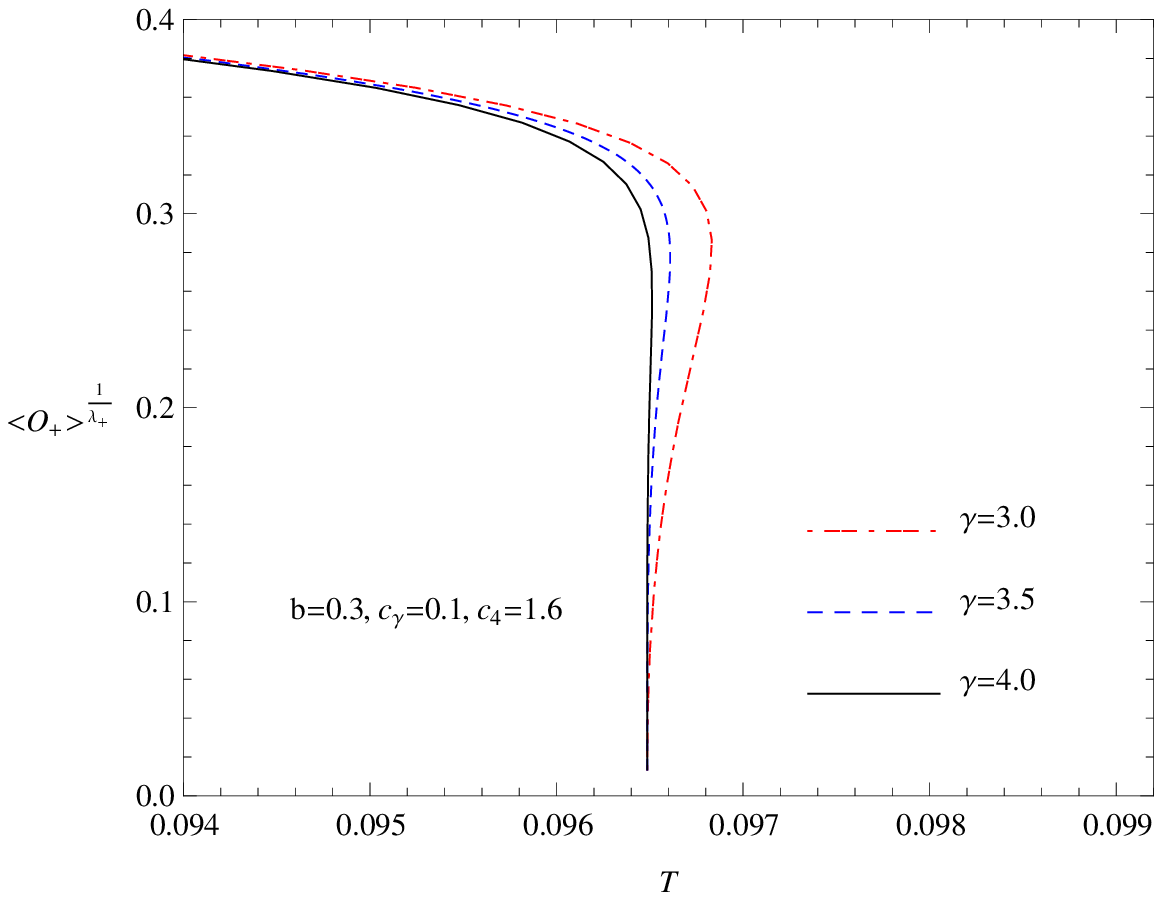}\;\;\;\;\includegraphics[width=7cm]{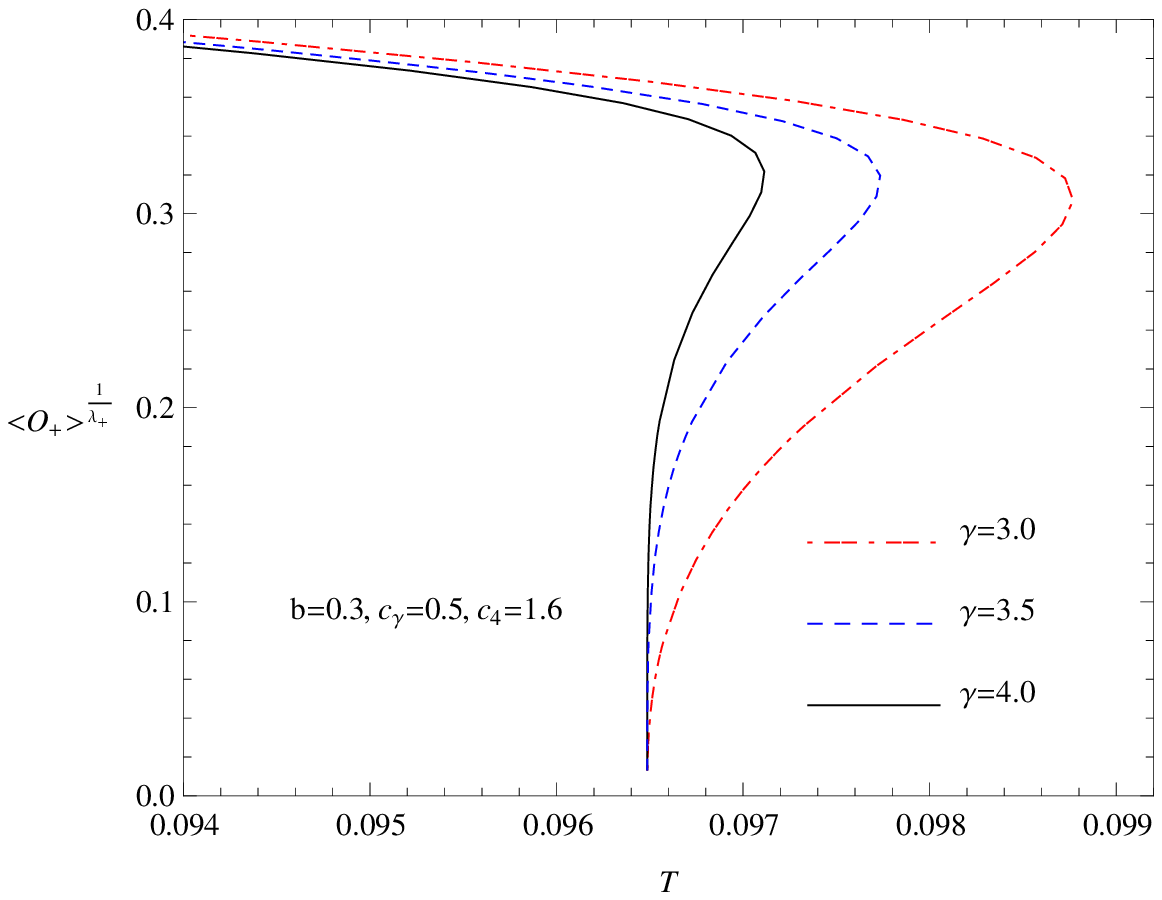}
\caption{The condensate
$\langle\mathcal{O_{+}}^{\frac{1}{\lambda_+}}\rangle$ as a function
of temperature with fixed values $c_{\gamma}$ for different values
of $b$ and  $\gamma$, which shows that the type of phase transition
depends on the coupling parameter $b$, model parameters $c_{\gamma}$
and $\gamma$. The three lines in each panel from left to right
correspond to decreasing $\gamma$, i.e, 4, 3.5 and 3.}
\end{center}
\end{figure}
coefficient $c_{\gamma}$ and decreases with the power $\gamma$ for
fixed the coupling parameter $b$, which is consistent with that
obtained in the case $b=0$. For fixed the model parameters
$c_{\gamma}$ and $\gamma$, we find that the presence of $b$ leads to
the critical temperatures for two types of phase transitions more
lower and makes the transition point of the phase transition from
the second order to the first order appears more hardly. Moreover,
we check the critical behaviors of the condensate
$\langle\mathcal{O_{+}}\rangle$ near the second-order phase
transition point with different values of $b$ and $\gamma$ for fixed
$c_{\gamma}=-1$ and $c_4=0.5$. From Fig. 3, we find the critical
exponent $\beta$ can be approximated as
\begin{figure}[ht]
\begin{center}
\includegraphics[width=5.1cm]{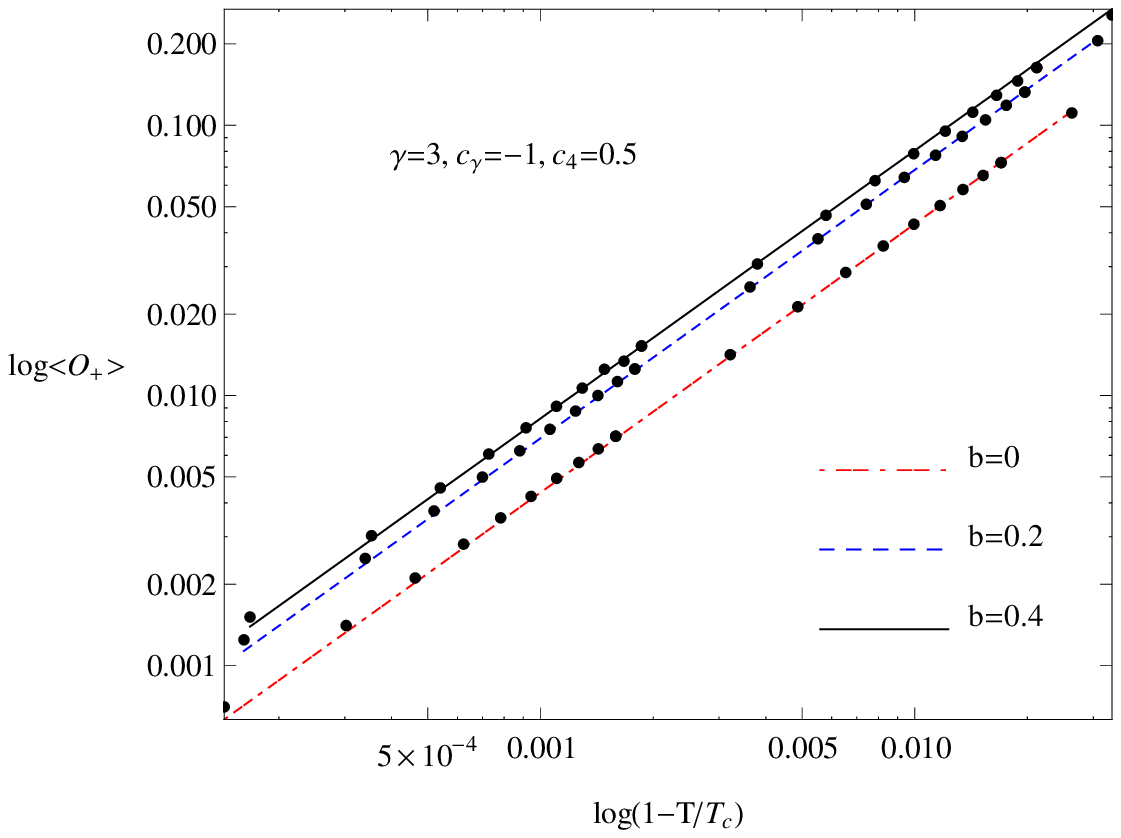}\;\;\;\;
\includegraphics[width=5.1cm]{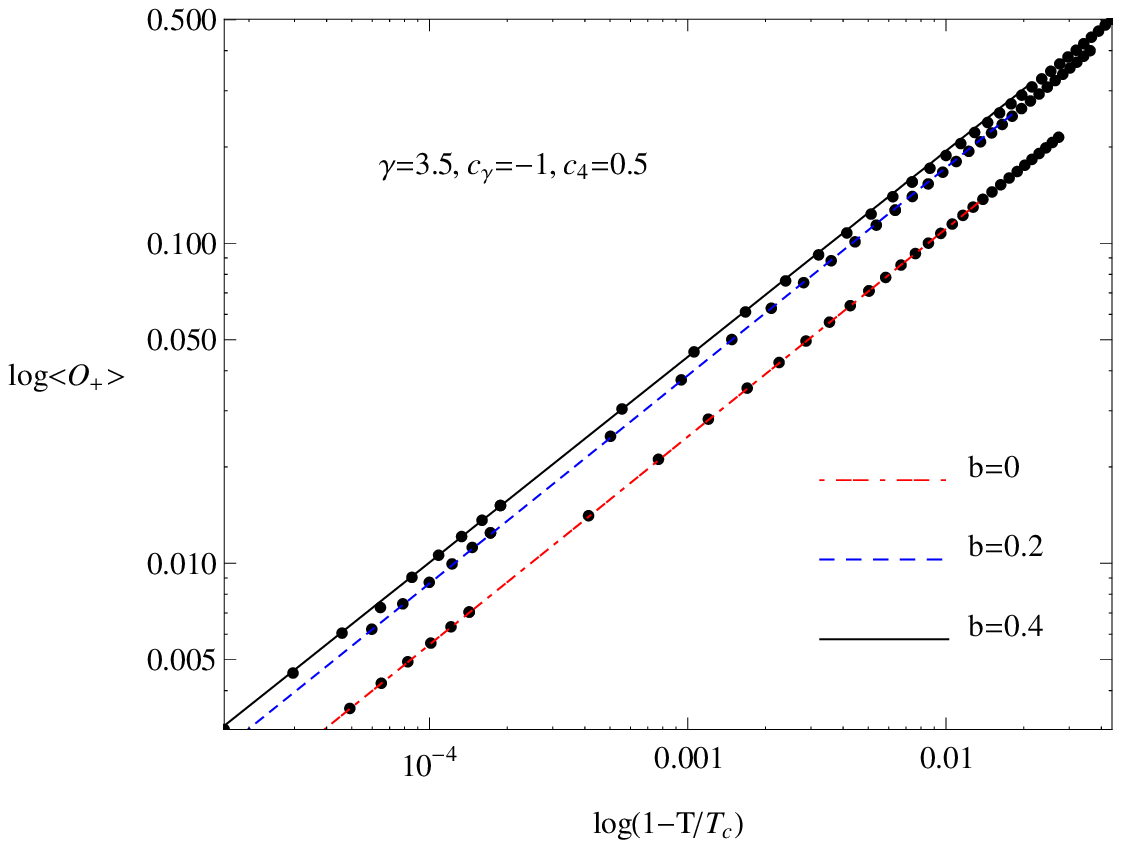}
\includegraphics[width=5.1cm]{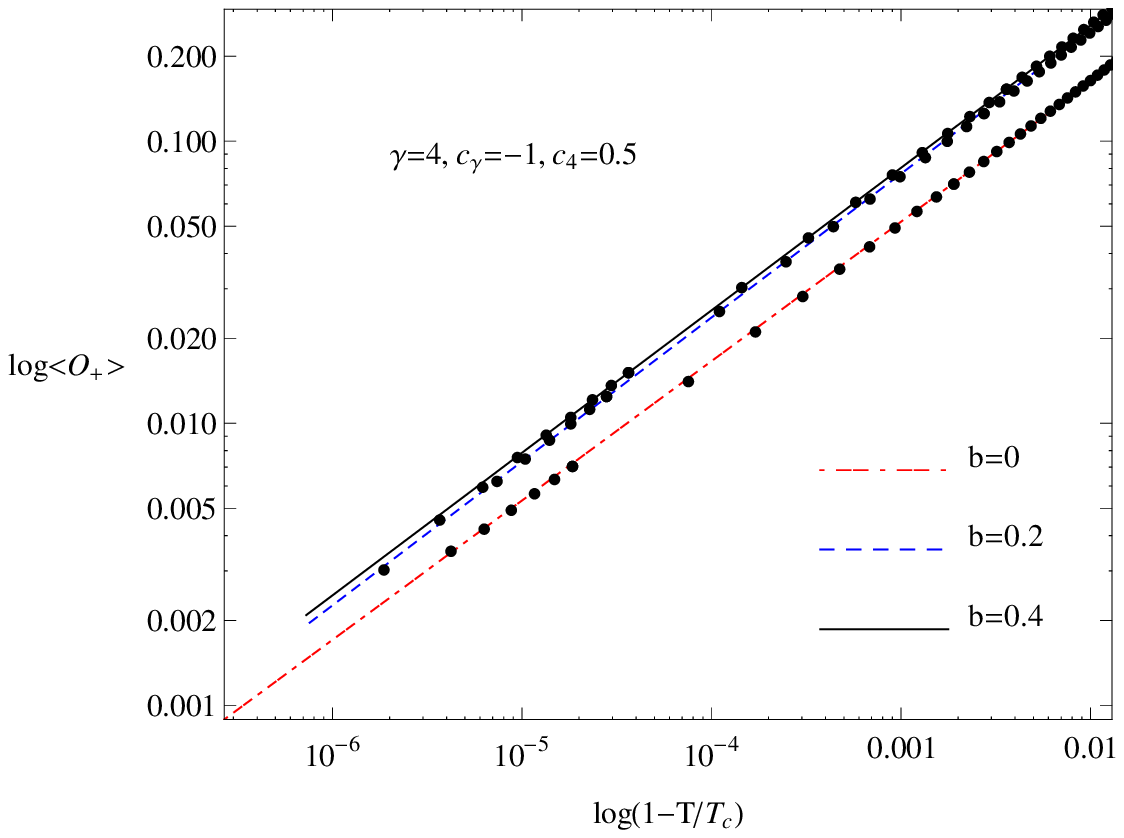}
\caption{The condensate $\langle\mathcal{O_{+}}\rangle$ vs $1-T/T_c$
in logarithmic scale with different values of $b$ and $\gamma$,
which shows that the slope is independent of $b$ but sensitive to
$\gamma$.}
\end{center}
\end{figure}
\begin{eqnarray}
\beta\simeq\frac{1}{\gamma-2}.
\end{eqnarray}
Obviously, the critical exponent $\beta$ is determined only by the
model parameters in the $\mathfrak{F}(\psi)$ and is independent of
the coupling parameter $b$, which implies that the critical exponent
$\beta$ does not depend on the scalar mass since the presence of the
coupling parameter $b$ changes the effective mass of the scalar
field. Combining with the results obtained in
\cite{a18,a19,a12,ap1}, we find that the dependence of  the critical
exponent $\beta$ near the second-order phase transition point only
on the model parameter $\gamma$ could be a universal property in
such a kind of general models for holographic superconductors.

\section{The electrical conductivity}

In this section we will investigate the electrical conductivity when
the scalar field coupling to Einstein's tensor and see what effects
of the coupling parameter $b$ and the model parameters
($c_{\gamma}$, $\gamma$ and $c_4$) on the electrical conductivity in
the planar Schwarzschild AdS black hole spacetime.

 Following the standard procedure in \cite{Hs01,Hs0,a1,Hs02,Hs03}, we suppose that the perturbation of the
vector potential is translational symmetric and has a time
dependence as $\delta A_x =A_x(r)e^{-i\omega t}$. In the probe
approximation, the effect of the perturbation of metric can be
ignored. Thus, equation of motion for $\delta A_x$ obeys to
\begin{eqnarray}
A_x^{''}+\frac{f'}{f}A_x'
+\bigg[\frac{\omega^2}{f^2}-(1+3b)\frac{\mathfrak{F}(\psi)}{f}\bigg]A_x=0.\label{de}
\end{eqnarray}
As $b$ tends to zero, the above equation can be reduced to that in
the case in which the scalar field is not coupling with Einstein's
tenor. Since there exists only the ingoing wave at the black hole
horizon, the boundary condition on $A_x$ near $r\sim r_H$ is $A_x=
f^{-\frac{i\omega }{3r_H}}$. In the asymptotic AdS region, one can
find easily that $A_x$ behaves like
\begin{eqnarray}
A_x=A^{(0)}_x+\frac{A^{(1)}_x}{r}.
\end{eqnarray}
According to the AdS/CFT, we can express the conductivity as
\begin{eqnarray}
\sigma(\omega)=\frac{\langle J_x\rangle}{E_x}=-\frac{i\langle
J_x\rangle}{\omega A_x}=-i\frac{A^{(1)}_x}{\omega A^{(0)}_x}.
\end{eqnarray}
\begin{figure}[ht]
\begin{center}
\includegraphics[width=5.1cm]{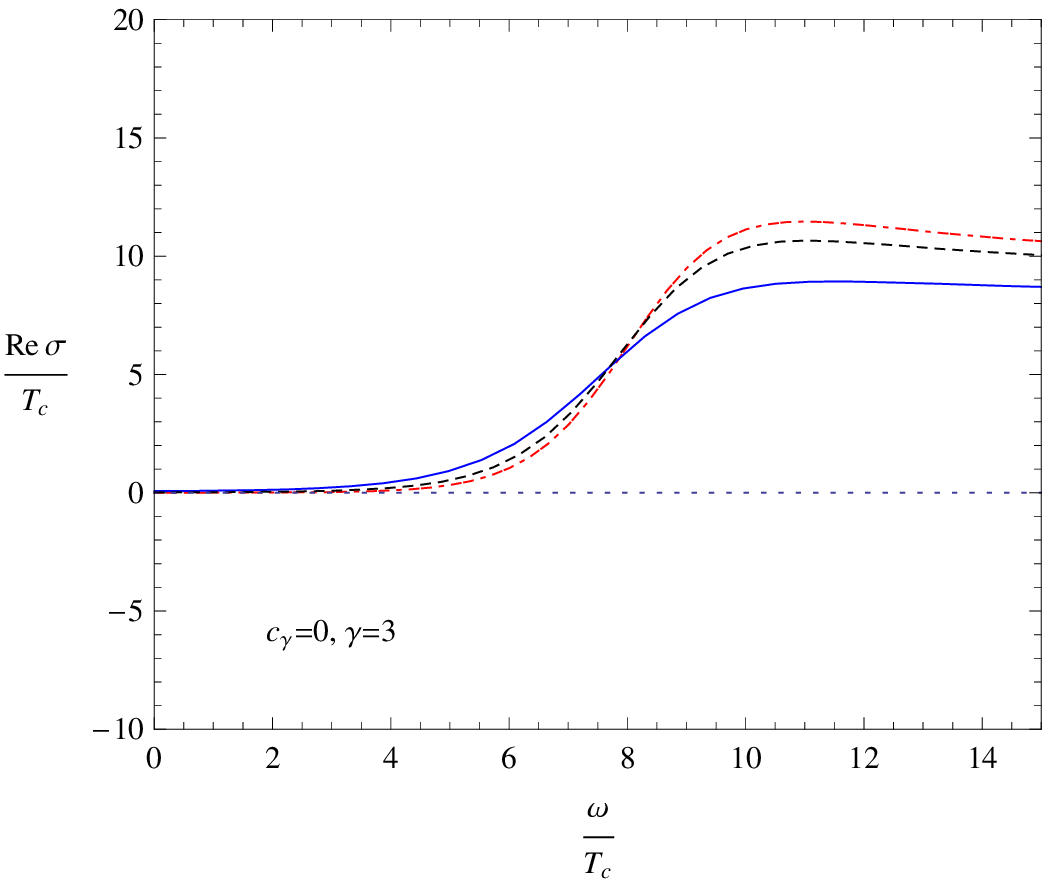}\;\;\;\;
\includegraphics[width=5.1cm]{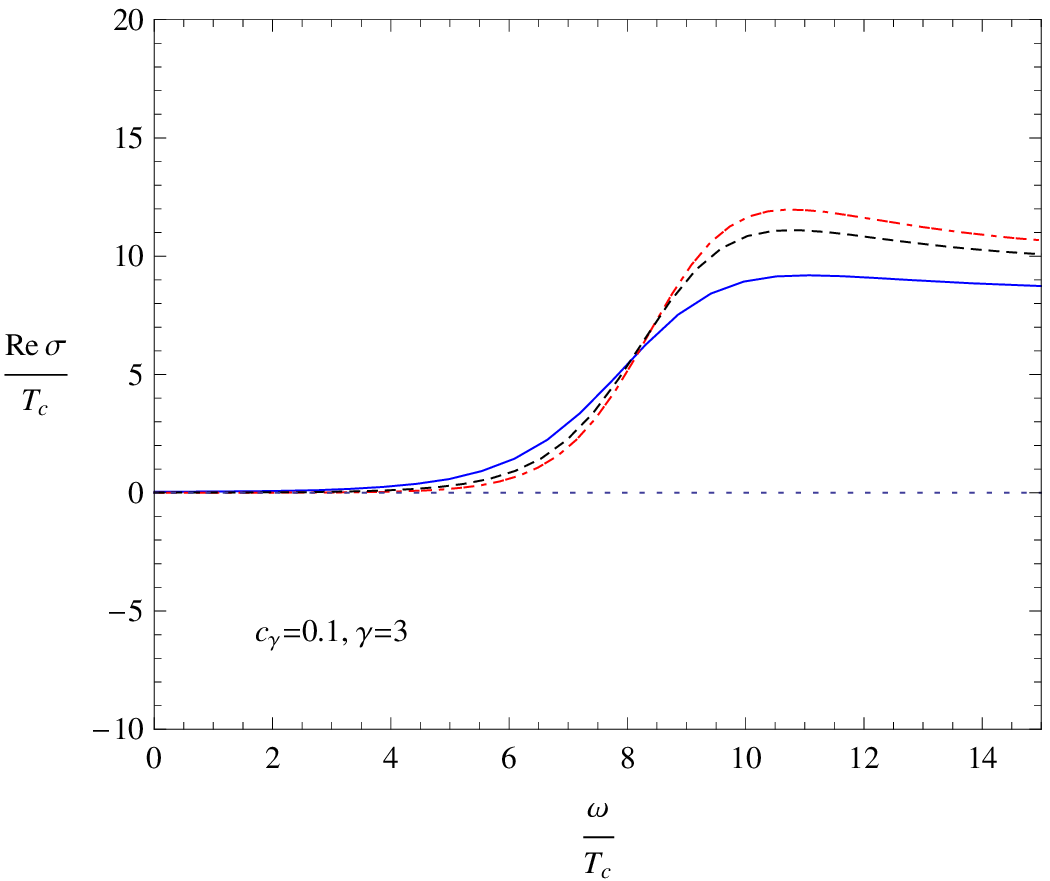}
\includegraphics[width=5.1cm]{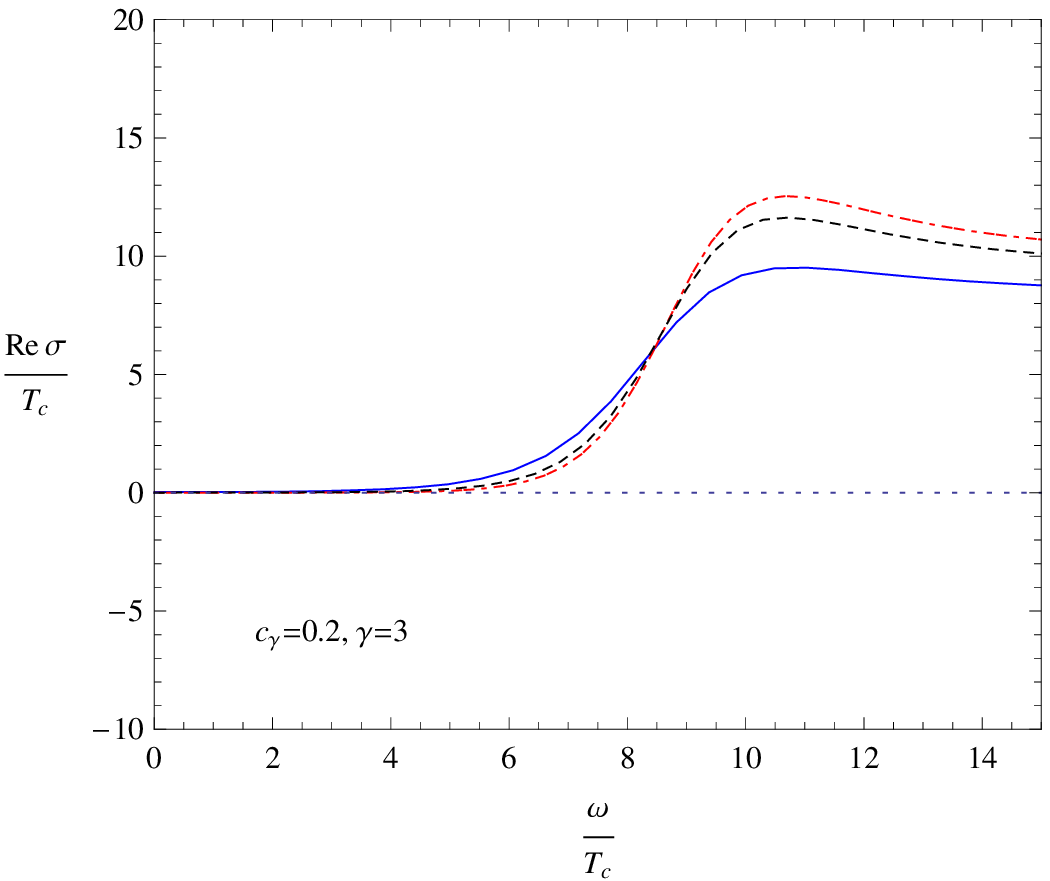}\\
\includegraphics[width=5.1cm]{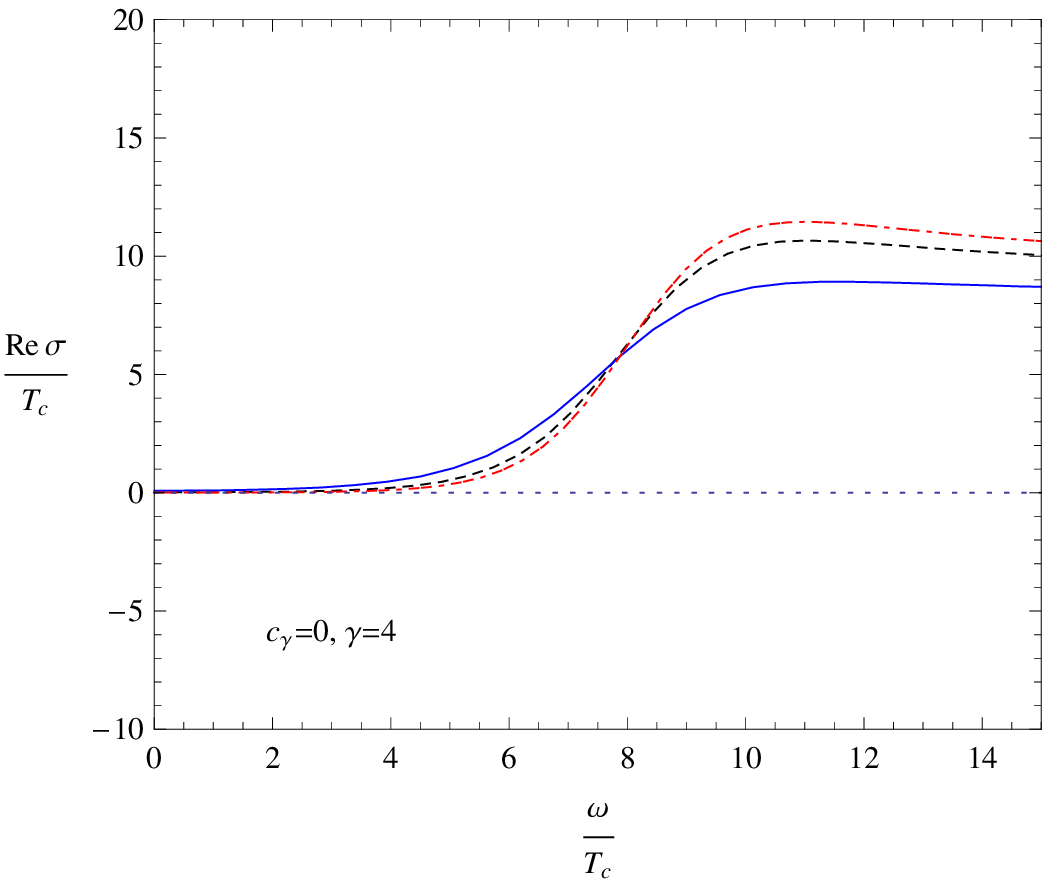}\;\;\;\;
\includegraphics[width=5.1cm]{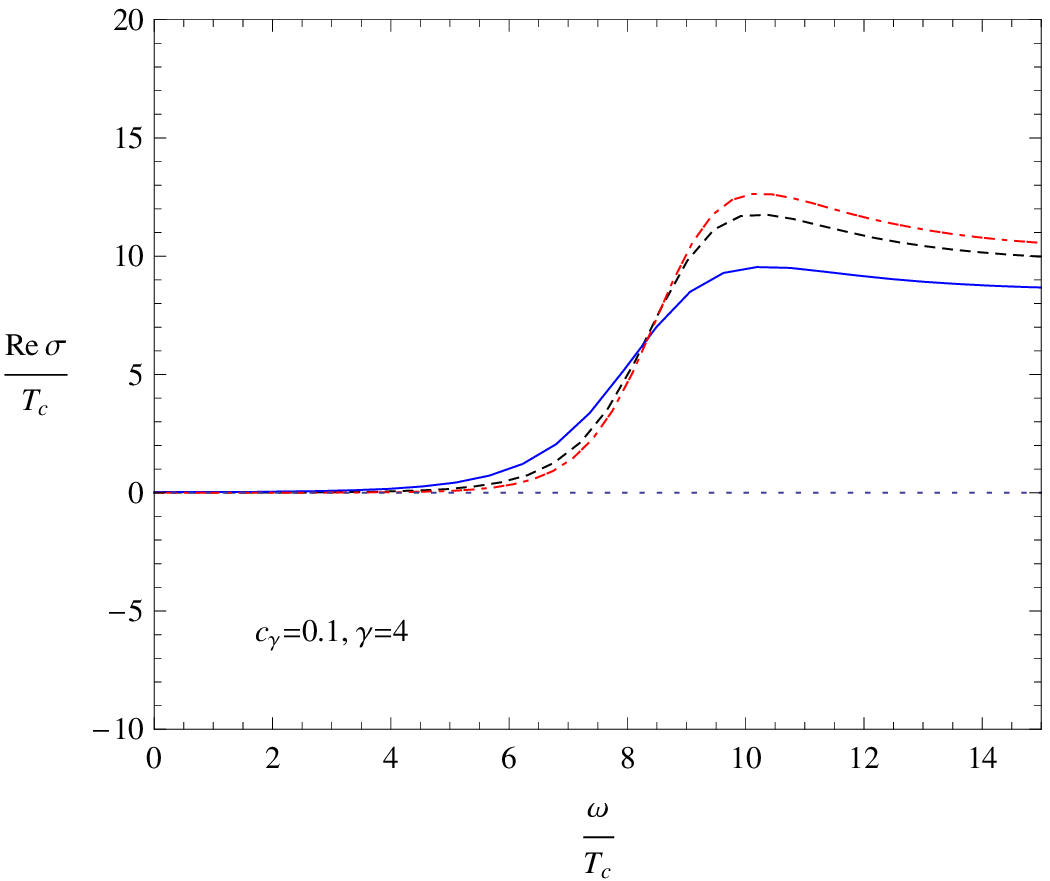}
\includegraphics[width=5.1cm]{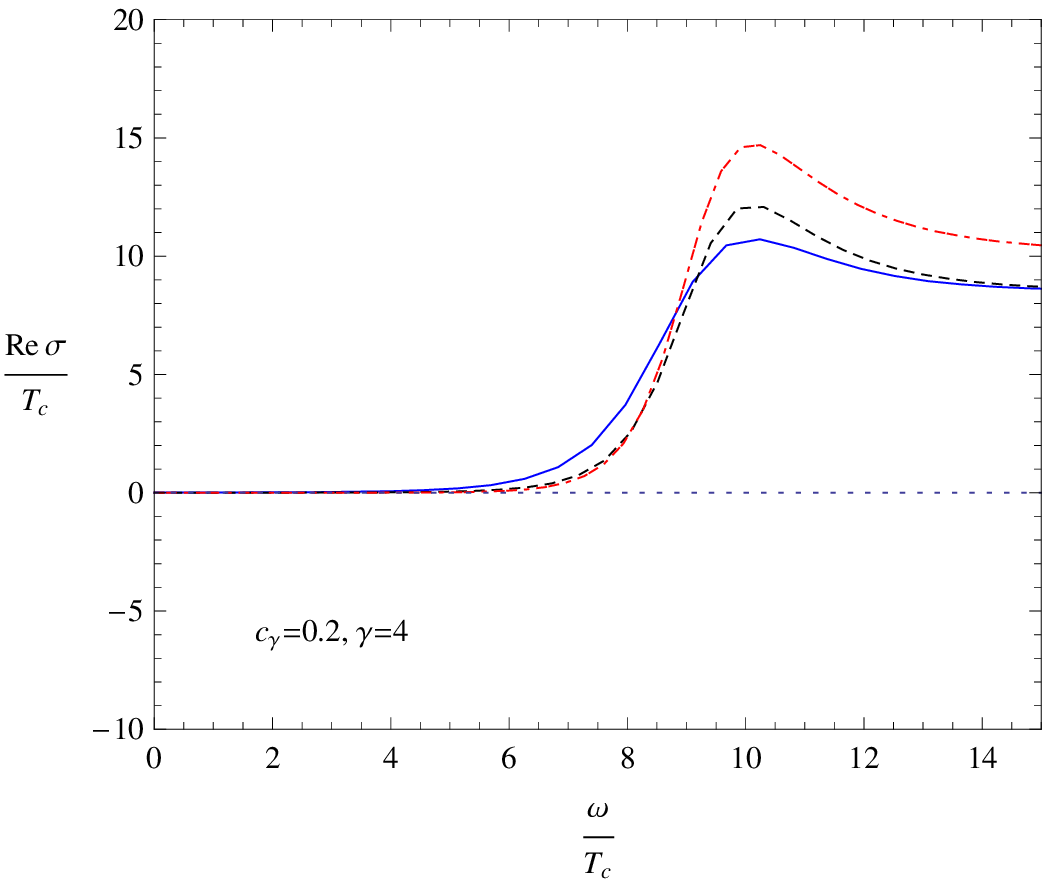}
\caption{The real part of of the conductivity with fixed values of
$b$ for different models with
$\mathfrak{F}(\psi)=\psi^2+c_{\gamma}\psi^{\gamma}$. The solid,
dashed and dash-dotted lines are corresponding to the cases with
$b=0$, $0.1$ and $0.2$, respectively.}
\end{center}
\end{figure}
\begin{figure}[ht]
\begin{center}
\includegraphics[width=5.1cm]{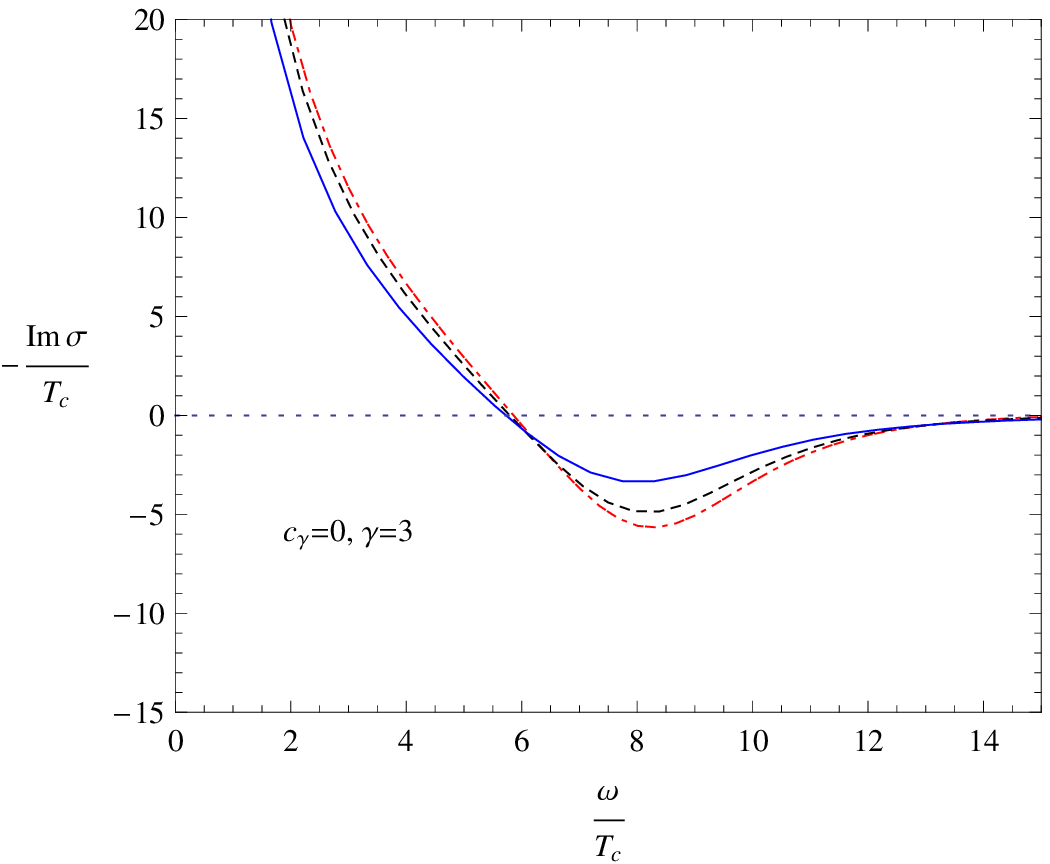}\;\;\;\;
\includegraphics[width=5.1cm]{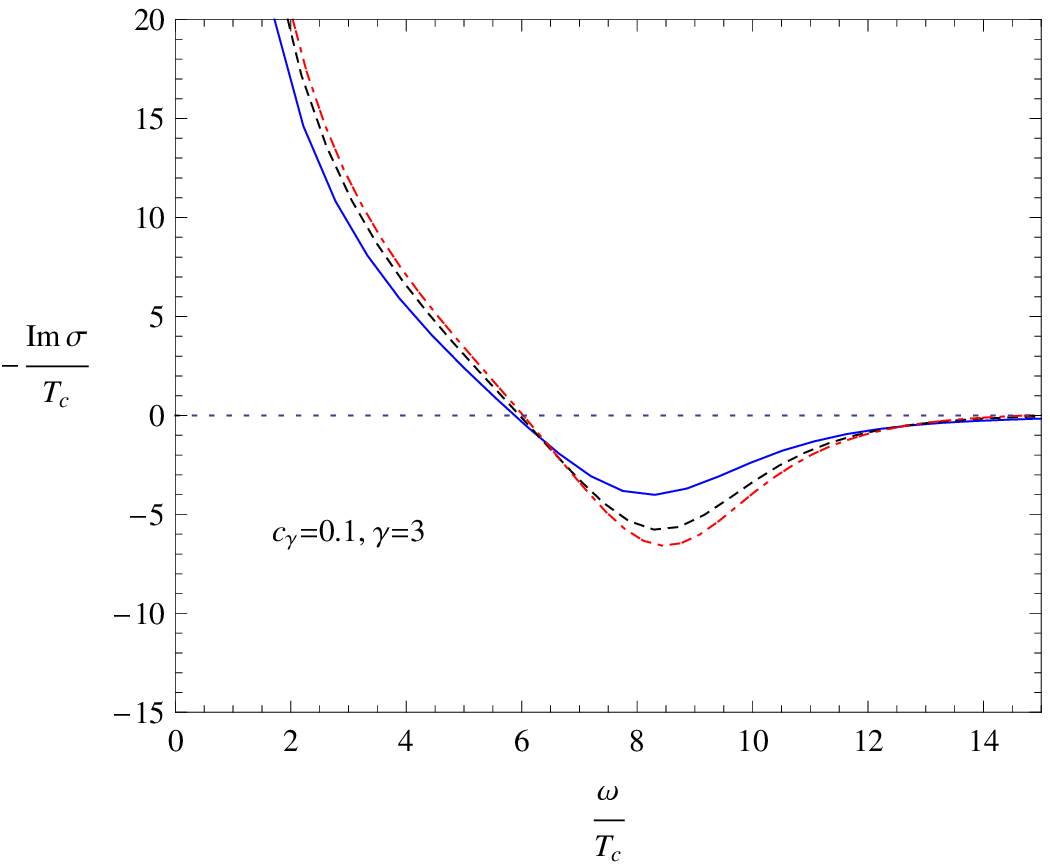}
\includegraphics[width=5.1cm]{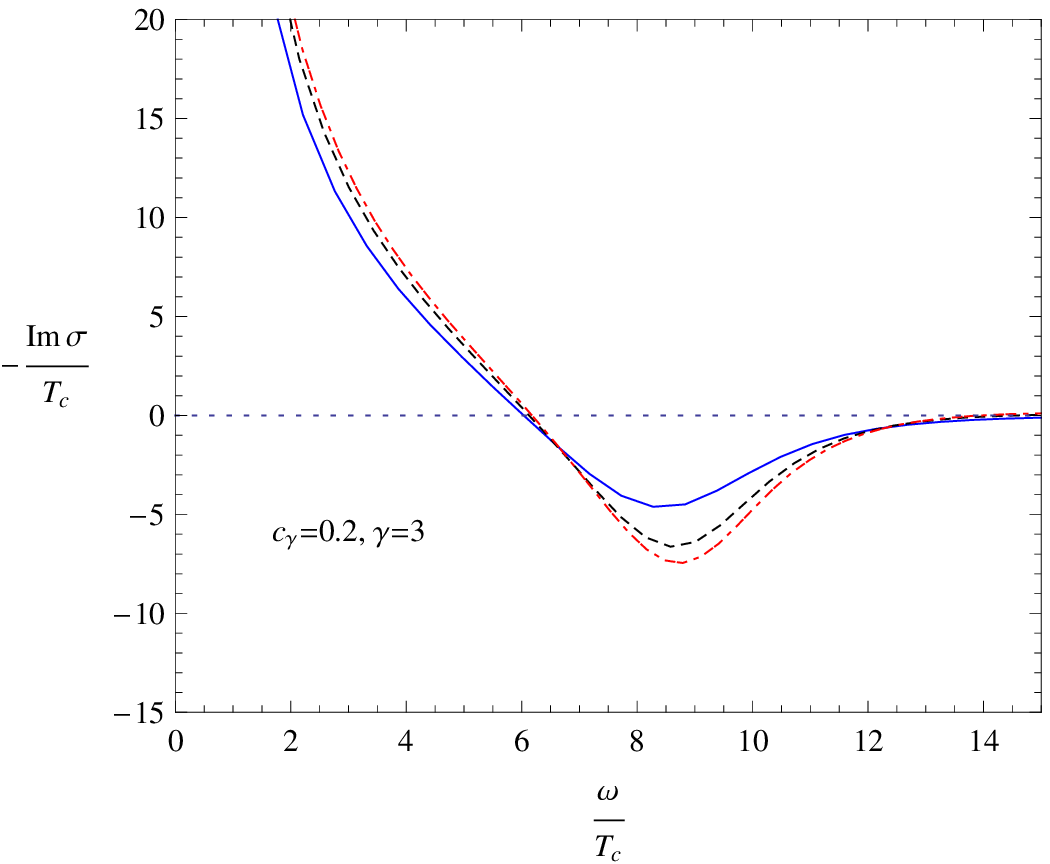}\\
\includegraphics[width=5.1cm]{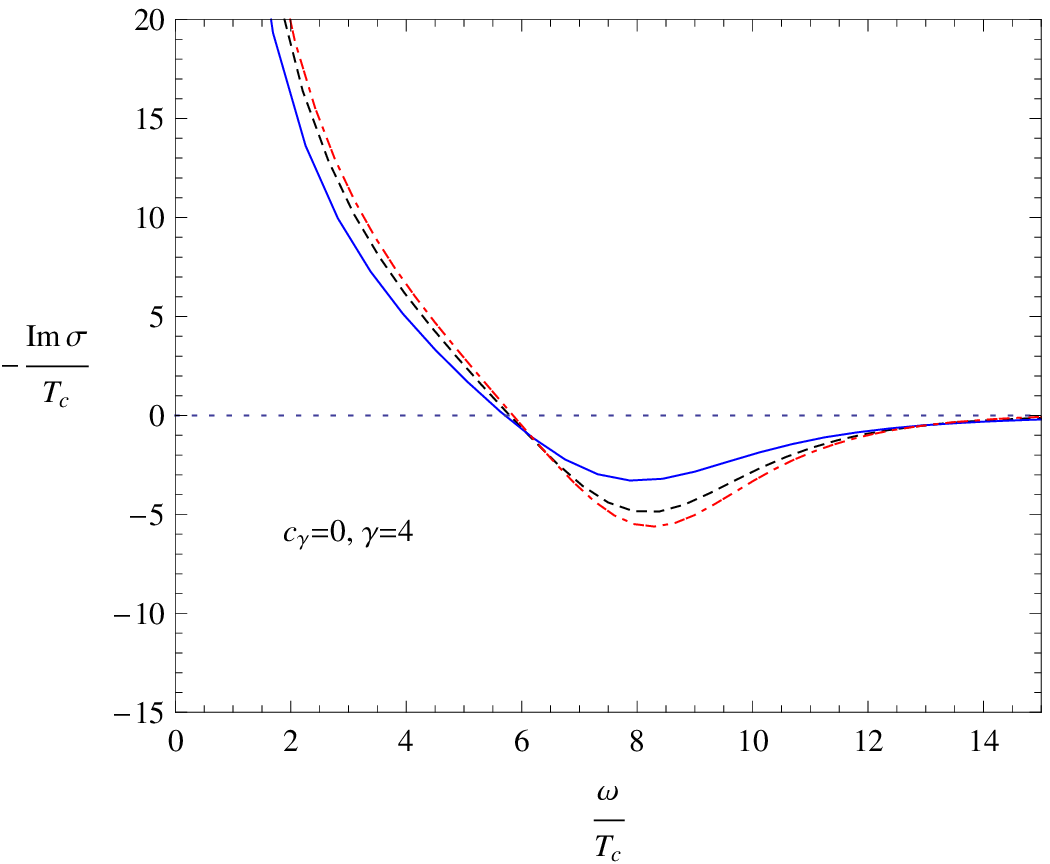}\;\;\;\;
\includegraphics[width=5.1cm]{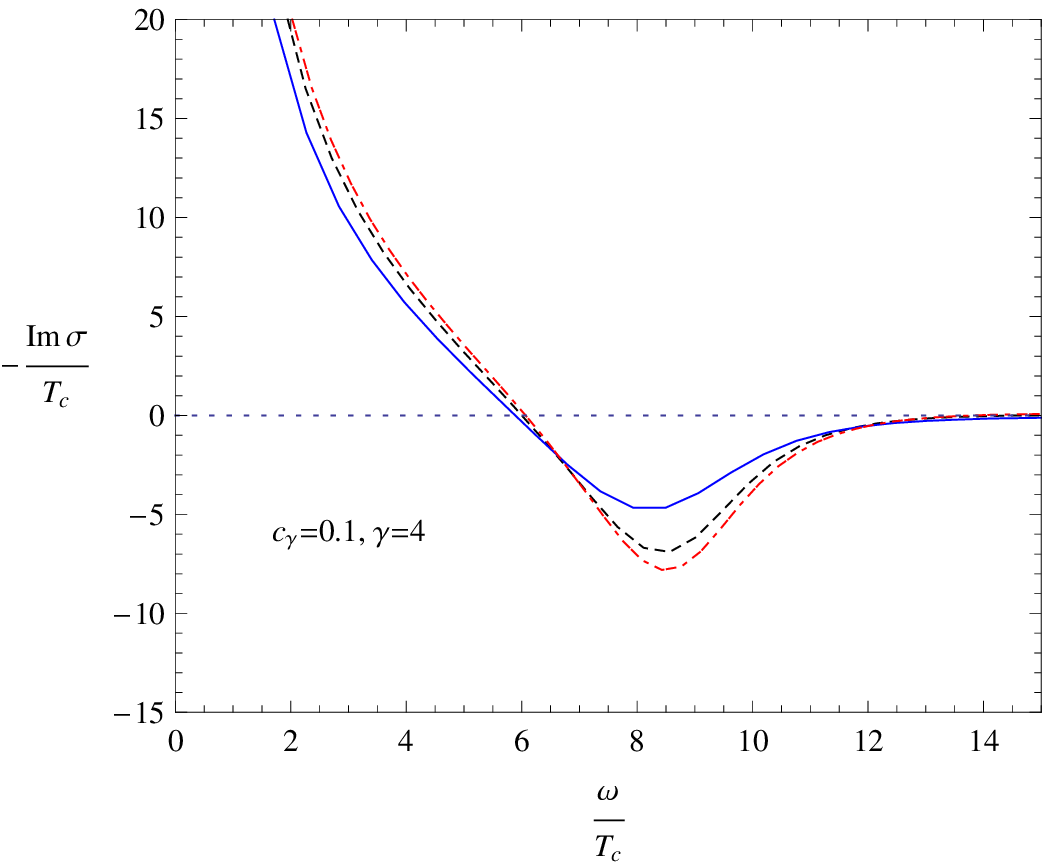}
\includegraphics[width=5.1cm]{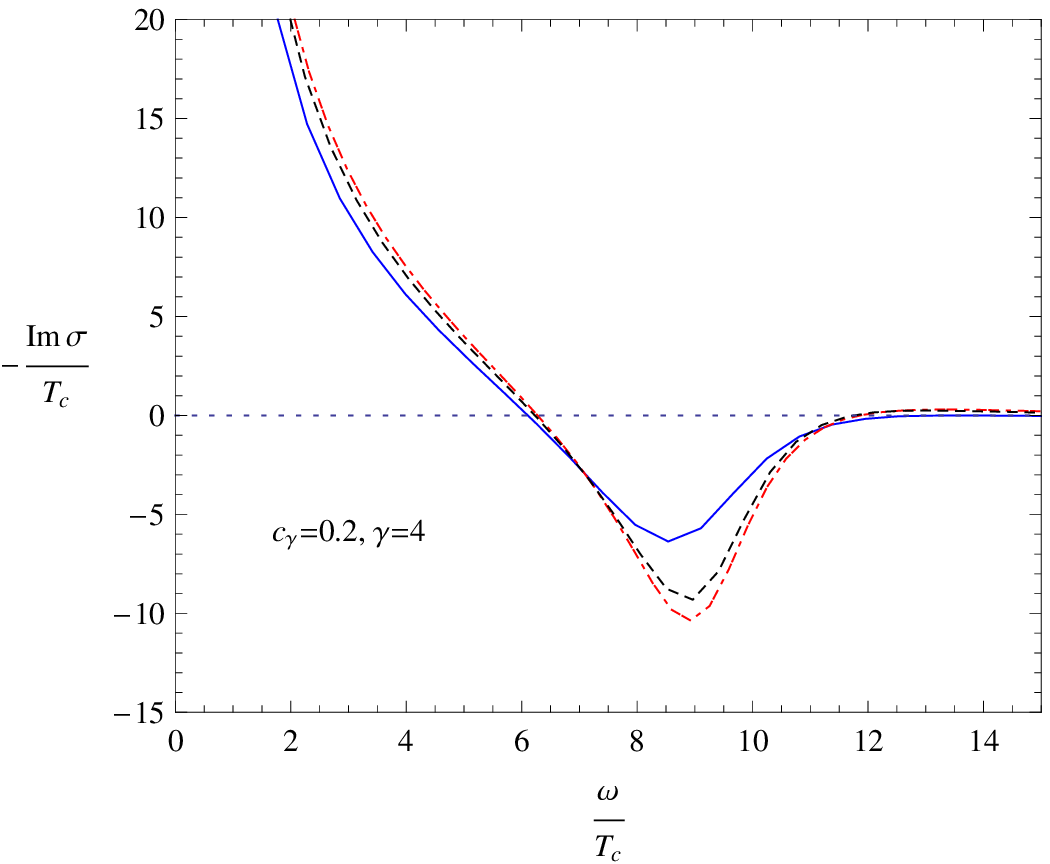}
\caption{The imaginary part of the conductivity with fixed values of
$b$ for different models with
$\mathfrak{F}(\psi)=\psi^2+c_{\gamma}\psi^{\gamma}$. The solid,
dashed and dash-dotted lines are corresponding to the cases with
$b=0$, $0.1$ and $0.2$, respectively.}
\end{center}
\end{figure}
Solving the motion equation (\ref{de}) numerically, we can obtain
the conductivity for the general forms of function
$\mathfrak{F}(\psi)=\psi^2+c_{\gamma}\psi^{\gamma}+c_4\psi^4$. Here
we set $c_4=0$ and $m^2L^2=-2$ for clarity.

In Figs. 4 and 5, we plot the frequency dependent conductivity for
operator $\langle\mathcal{O}_{+}\rangle$ with different values of
$b$, $c_{\gamma}$ and $\gamma$. Fixing the coupling parameter $b$,
the gap frequency $\omega_g$ increase with $c_{\gamma}$ for fixed
$\gamma$, grows  with $\gamma$ for fixed $c_{\gamma}$. These
behaviors of the gap frequency $\omega_g$ are also observed in the
case $b=0$. For fixed model parameter $c_{\gamma}$ and $\gamma$, we
find that the gap frequency $\omega_g$ becomes bigger for the larger
$b$. It implies that the presence of the coupling parameter $b$
enhances the strength of the strong coupling in holographic
superconductors. Our results show that not only the form of the
scalar field $\mathfrak{F}(\psi)$, but also the coupling between the
scalar field and Einstein's tensor of the background will affect the
so-called universal relation $\omega_g/T_c\simeq 8$.

\section{Summary}

In this paper we studied a general class of the holographic
superconductor models via the St\"{u}ckelberg mechanism in the
non-minimal derivative coupling theory in which the charged scalar
field is kinetically coupling with Einstein's tensor.  In the probe
limit, we found that the coupling parameter $b$ decreases the
critical temperature and makes the formation of the scalar
condensate more hardly. Our results also show that both the coupling
parameter and the parameters which define $\mathfrak{F}(\psi)$ can
separate the first and second order phase transitions. With the
increase of $b$ , the transition point of the phase transition from
the second order to the first order also appears hardly in this
generalized system. This means that the influences of $b$ on the
phase transition are different from that of the Gauss-Bonnet
coupling parameter $\alpha$. For the second order phase transition,
we observed that the deviation of the critical exponents from that
of the mean field result is independent of the coupling parameter
$b$ and is determined only by the model parameters in the
$\mathfrak{F}(\psi)$.

We also calculated the electric conductivity numerically and found
that the gap frequency $\omega_g$ depends not only on the form of
the scalar field $\mathfrak{F}(\psi)$, but also on the coupling
between the scalar field and Einstein's tensor of the background.
With increase of the coupling parameter $b$, the ratio of the gap
frequency in conductivity $\omega_g$ to the critical temperature
$T_c$ increases, which implies that the coupling between the scalar
field and Einstein's tensor enhances the strength of the strong
coupling in holographic superconductors.

\begin{acknowledgments}
This work was partially supported by the National Natural Science
Foundation of China under Grant No.10875041, the construct program
of key disciplines in Hunan Province and the Project of Knowledge
Innovation Program (PKIP) of Chinese Academy of Sciences under Grant
No. KJCX2.YW.W10. J. Jing's work was partially supported by the
National Natural Science Foundation of China under Grant
No.10875040, No.10935013 and 973 Program Grant No. 2010CB833004.
\end{acknowledgments}


\vspace*{0.2cm}
\begin{thebibliography}{99}
\baselineskip=0.6 cm

\bibitem{ads1} J. M. Maldacena,  Adv. Theor. Math. Phys. {\bf2}, 231
 (1998), [hep-th/9711200].


\bibitem{ads2} E. Witten,  Adv. Theor. Math. Phys. {\bf 2}, 253 (1998).

\bibitem{ads3} S. S. Gubser, I. R. Klebanov, and A. M. Polyakov,  Phys. Lett. B {\bf 428},
105 (1998).

\bibitem{Hs01} S. S. Gubser,  Phys. Rev. D {\bf78}, 065034 (2008).

\bibitem{Hs0} S. A. Hartnoll, C. P. Herzog and G. T. Horowitz,  Phys. Rev. Lett. {\bf101}, 031601
(2008).

\bibitem{a1} S. A. Hartnoll, C. P. Herzog and G. T. Horowitz, J. High Energy Phys. {\bf 0812}, 015 (2008).

\bibitem{Hs02} S.A. Hartnoll, Class. Quant. Grav. {\bf26}, 224002 (2009), [arXiv:0903.3246].

\bibitem{Hs03} C. P. Herzog,  J. Phys. A {\bf 42}, 343001 (2009).

\bibitem{Hsf0} P. Basu, A. Mukherjee and H. H. Shieh, Phys. Rev. D {\bf 79}, 045010 (2009).

\bibitem{Hsf01}C. P. Herzog, P. K. Kovtun, and D. T. Son,  Phys. Rev. D {\bf79},
066002 (2009).

\bibitem{Hsa1} S. S. Gubser, Class. Quant. Grav. {\bf 22}, 5121 (2005).

\bibitem{a0} R. Gregory, S. Kanno, and J. Soda, J. High Energy Phys. {\bf0910}, 010
(2009).

\bibitem{a01}H. Zeng, Z. Fan, Z. Ren, Phys. Rev. D {\bf 80}, 066001 (2009).


\bibitem{a2} S. S. Gubser, Phys. Rev. Lett.{\bf 101}, 191601 (2008).

\bibitem{a3} M. M. Roberts and S. A. Hartnoll, J. High Energy Phys. {\bf 0808}, 035
(2008).

\bibitem{a4} S. S. Gubser and S. S. Pufu, J. High Energy Phys. {\bf 0811}, 033 (2008).


\bibitem{a40}  R. Cai and H. Zhang,  Phys. Rev. D {\bf81}, 066003
(2010), arXiv:0911.4867.

\bibitem{a401}  Q. Y. Pan, B. Wang, E. Papantonopoulos, J. Oliveira and A.
Pavan,  Phys. Rev. D {\bf81}, 106007 (2010).

\bibitem{a402} J. Jing, L. Wang, S. Chen, arXiv:1001.2946.


\bibitem{a41} S. Sin, S. Xu and Y. Zhou, arXiv:0909.4857.

\bibitem{a42} E. J. Brynjolfsson, U.H. Danielsson, L. Thorlacius and T. Zingg,
J. Phys. A {\bf43}, 065401 (2010). [arXiv:0908.2611]


\bibitem{a5} F. Denef and S. A. Hartnoll,  arXiv:0901.1160.

\bibitem{a6} S. S. Gubser, C. P. Herzog, S.S. Pufu and T. Tesileanu,
 Phys. Rev. Lett. {\bf 103}, 141601 (2009)

\bibitem{a7} J. P. Gauntlett, J. Sonner and T. Wiseman, Phys. Rev. Lett. {\bf 103},151601
(2009).[arXiv:0907.3796].


\bibitem{a8} S. S. Gubser, S. S. Pufu and F. D. Rocha, Phys. Lett. B {\bf683},201 (2010).

\bibitem{a8d1} M. Ammon, J. Erdmenger, M. Kaminski and P. Kerner,
Phys. Lett. B {\bf 680} 516 (2009);

M. Ammon, J. Erdmenger, M. Kaminski and P. Kerner, J. High Energy
Phys. {\bf0910} 067 (2009).

\bibitem{GT}G. T. Horowitz and M. M. Roberts, Phys. Rev. D {\bf78}, 126008 (2008).

\bibitem{a81} S. S. Gubser and A. Nellore, arXiv:0908.1972.

\bibitem{a9} G. T. Horowitz and M. M. Roberts, J. High Energy Phys. {\bf 0911}, 015
(2009). arXiv:0908.3677.

\bibitem{a10} R. A. Konoplya and A. Zhidenko, Phys. Lett. B {\bf686}, 199, (2010), arXiv: 0909.2138.

\bibitem{a100} T. Nishioka, S. Ryu and T. Takayanagi,  J. High Energy Phys. {\bf1003},131 (2010), arXiv:
0911.0962.


\bibitem{a20} X. H. Ge, B. Wang, S. F. Wu, and G. H. Yang, arXiv:1002.4901 [hep-th].

\bibitem{a21} Y. Brihaye and B. Hartmann, Phys. Rev. D 81, 126008 (2010), arXiv:1003.5130.

\bibitem{a22} L. Barclay, R. Gregory, S. Kanno, and P. Sutcliffe,  J. High Energy Phys.
{\bf1012},029(2010),  arXiv:1009.1991[hep-th].

\bibitem{a23} Rong-Gen Cai, Zhang-Yu Nie, and Hai-Qing Zhang, Phys. Rev. D {\bf82},
066007 (2010); arXiv:1007.3321.

\bibitem{a13} P. Basu, J. He, A. Mukherjee and H. Shieh, Phys. Lett. B {\bf689}, 45 (2010), arXiv: 0911.4999.


\bibitem{a14} J. Sonner, Phys. Rev. D {\bf 80}, 084031 (2009).

\bibitem{a15} E. Nakano1 and W. Wen, Phys. Rev. D {\bf 78}, 046004
(2008).

\bibitem{a151}  S. Chen, L. Wang, C. Ding and J. Jing, Nucl. Phys. B {\bf836}, 222(2010),
arXiv:0912.2397.

\bibitem{a152} J. Jing and S. Chen, Phys. Lett. B {\bf686}, 68 (2010)

\bibitem{a16s}  M. Ammon, J. Erdmenger, V. Grass, P. Kerner and A.
O'Bannon, Phys. Lett. B {\bf686},192 (2010), arXiv: 0912.3515.

\bibitem{a17} K. Maeda, M. Natsuume and T. Okamura, Phys. Rev. D{\bf79},126004
(2009);

K. Maeda, M. Natsuume and T. Okamura, Phys. Rev. D{\bf81}, 026002
(2010).


\bibitem{a18} S. Franco, A.M. Garcia-Garcia, and D. Rodriguez-Gomez, J. High
Energy Phys. {\bf04}, 092 (2010).

\bibitem{a19} S. Franco, A.M. Garcia-Garcia, and D. Rodriguez-Gomez, Phys.
Rev. D {\bf81}, 041901(R) (2010).


\bibitem{ap1} Q. Y. Pan, B. Wang, Phys. Lett. B {\bf693}, 159 (2010).

\bibitem{ap2}J. Jing, L. Wang, Q. Y. Pan, and S. Chen, arXiv:
1012.0644.


\bibitem{a12} F. Aprile and J. Russo, Phys. Rev. D {\bf81}, 026009 (2010).

\bibitem{Sushkov:2009} S. V. Sushkov, Phys. Rev. D {\bf 80}, 103505  (2009).

\bibitem{Gao:2010} C. J. Gao, JCAP {\bf06}, 023 (2010), arXiv: 1002.4035.

\bibitem{Granda:2009} L.N. Granda, arXiv: 0911.3702.

\bibitem{Saridakis:2010} E. N. Saridakis and S. V. Sushkov, Phys. Rev. D {\bf 81}, 083510 (2010), arXiv: 1002.3478.

\bibitem{sc1} S. Chen and J. Jing, Phys. Rev. D {\bf82},
084006(2010).

\bibitem{Chen:2010} S. Chen and J. Jing, Phys. Lett. B {\bf691}, 254
(2010), arXiv: 1005.5601.

\bibitem{sc2} C. Ding, C. Liu, J. Jing and S. Chen, J. High
Energy Phys. {\bf1011}, 146 (2010).

\bibitem{ads} I. R. Klebanov and E. Witten,  Nucl. Phys. B {\bf556}, 89 (1999).
\end{thebibliography}
\end{document}